\documentstyle[12pt,preprint,revtex]{aps}
\begin{document}
\def\overlay#1#2{\setbox0=\hbox{#1}\setbox1=\hbox to \wd0{\hss #2\hss}#1%
\hskip -2\wd0\copy1}
\rightline{{\sl JETP Letters},\ {\bf 58},\ No.\ 11,\ (1993)}
\begin{title}
\begin {center}
{TEMPERATURE DEPENDENCE OF THE UPPER CRITICAL FIELD IN HIGH-TEMPERATURE
SUPERCONDUCTORS:\ LOCALIZATION EFFECTS.}
\end{center}
\end{title}
\author{E.Z.Kuchinskii,\ M.V.Sadovskii}
\begin{instit}
Institute for Electrophysics,\ Russian Academy of Sciences,\ Ural Branch,\\
Ekaterinburg,\ 620219,\ Russia
\end{instit}
\begin{abstract}

It is shown that the anomalous temperature dependence of the orbital part of
the upper critical field $H_{c2}$ observed for epitaxially grown films of
high-temperature superconductor $Bi-Sr-Cu-O$ (in wide temperature interval)
can be satisfactorily explained by the influence of localization effects in
two-dimensional (quasi-two-dimensional) case.

\end{abstract}

\vskip 0.5cm

In a recent paper Osofsky et al. [1] presented the unique data on the
temperature dependence of the upper critical field of high-temperature
superconductor $Bi_{2}Sr_{2}CuO_{y}$ in wide temperature interval from
$T_{c}\approx 19K$ to $T\approx 0.005T_{c}$,\ which has shown rather anomalous
dependence with negative curvature at any temperature. The authors of [1]
has noted that this type of behavior is difficult to explain within any known
theory. It is sharply different from the standard behavior of BCS-model,\ as
well as from that predicted by the model of compact charged Bosons
(bipolarons) [2]. In this latter case $H_{c2}$ there appears a power-like
divergence for $T\rightarrow 0$, while the experimental value of $H_{c2}(T=0)$
is apparently finite. The aim of the present work is to demonstrate that the
observed dependence of $H_{c2}(T)$ can be satisfactorily explained by
localization effects in two-dimensional (quasi-two-dimensional) model in the
limit of sufficiently strong disorder [3]. Measurements of $H_{c2}$
in Ref. [1] were performed on epitaxially grown films of $Bi_{2}Sr_{2}CuO_{y}$,
however it is quite possible that the films were still disordered enough,
which can be guessed from rather wide ($\sim 7K$) superconducting transition.
Unfortunately the relevant data, in particular on the value of conductivity of
the films studied are absent. This gives us some grounds to try to interpret
the data obtained in Ref. [1] in the framework of rather strong disorder the
effects of which are obviously enhanced by the quasi-two-dimensional nature of
high-temperature superconductors.

Below we shall limit ourselves by description of purely two-dimensional case
because the appropriate dependences for quasi-two-dimensional system differ
only
very slightly for the relevant values of parameters of the model [3]. The
general approach to $H_{c2}$-behavior in strongly disordered systems was given
in Ref. [4].

To describe the electronic properties of strongly disordered system in external
magnetic field we need following Matsubara-type two-particle Green's functions
in diffusion and Cooper channels for small $q$ and $\omega_{m}$:
\begin{equation} \left\{ \begin{array}{l} \Phi_{E}({\bf
q}\omega_{m}=2\varepsilon_{n})\\ \Psi_{E}({\bf q}\omega_{m}=2\varepsilon_{n})
\end{array} \right\} =-\frac{N(E)}{i|\omega_{m}|+i\left\{\begin{array}{l}
D_{1}(\omega_{m})\\ D_{2}(\omega_{m}) \end{array}\right\} q^{2}} \label{Grin1}
\end{equation}
Here $\omega_{m}=2\pi mT$, $\varepsilon_{n}=2\pi
(n+\frac{1}{2})T$---are Matsubara frequencies, $N(E)$---the electronic density
of states at the Fermi level $E$.
The generalized "diffusion coefficients" $D_{1}$ и $D_{2}$ are in general
unequal in the presence of magnetic field which breaks time-invariance.
In this case we have to consider the coupled system of equations for both
Green's functions.

In the following we shall be interested only in the case of magnetic field
perpendicular to the highly conducting planes. From the standard approach to
superconducting transition in external magnetic field [5] we obtain the
following equation for temperature dependence of $H_{c2}(T)$:
\begin{equation}
ln\frac{T}{T_{c}}=2\pi
T\sum_{\varepsilon_{n}}\left\{ \frac{1}{2|\varepsilon_{n}|+2\pi
D_{2}(2|\varepsilon_{n}|)\frac{H}{\Phi_{0}}}-\frac{1}{2|\varepsilon_{n}|}
\right\}
\label{Hc}
\end{equation}
where $\Phi_{0}=\frac{\pi c}{e}$---is magnetic flux quantum,
$T_{c}$---is BCS transition temperature in the absence of magnetic field.

It is seen from Eq.(\ref{Hc}) that the anomalies in behavior of the upper
critical field are related to the frequency dependence of diffusion coefficient
which becomes non-trivial close to the Anderson metal-insulator transition.

Within the approach based upon self-consistent theory of localization [6,7]
the system of equations for diffusion coefficients in magnetic field in two-
dimensional case takes the following form [8]:
\begin{eqnarray}
\frac{D_{0}}{D_{2}}=1+\frac{1}{\pi
N(E)}\sum_{|{\bf q}| < q_{0}}\frac{1}{\omega+D_{1}q^{2}} \nonumber\\
\frac{D_{0}}{D_{1}}=1+\frac{1}{\pi N(E)}\sum_{|{\bf k}| <
q_{0}}\frac{1}{\omega+D_{2}k^{2}} \label{sisdif}
\end{eqnarray}
where $k^{2}=4m\omega_{H}(n+\frac{1}{2})$, $\omega_{H}=\frac{eH}{mc}$---
is the cyclotron frequency, $n$---is Landau's quantum number, $q_{0}$---
is the cut-off momentum ($q_{0}\approx l^{-1}$), defined by
$D_{0}q_{0}^{2}=\frac{1}{2\tau}$, where $D_{0}$---is the usual Drude diffusion
coefficient, $\tau^{-1}$---is the mean free time and $l$---is the mean free
path.

Let us introduce $d_{1}=\frac{D_{1}}{D_{0}}$, $d_{2}=\frac{D_{2}}{D_{0}}$
and dimensionless disorder parameter $\lambda=\frac{1}{2\pi E\tau}$. Then
Eqs.(\ref{sisdif}) are rewritten as:
\begin{eqnarray}
\frac{1}{d_{2}}&=&1+\frac{\lambda}{d_{1}}ln(1+d_{1}\frac{1}{2 \omega\tau})
\nonumber\\ \frac{1}{d_{1}}&=&1+\frac{\lambda}{d_{2}}\sum_{n=0}^{N_{0}}
\frac{1}{n+\frac{1}{2}+\frac{\omega}{4m\omega_{H}D_{0}} \frac{1}{d_{2}}}
\label{sisdif1}
\end{eqnarray}
where $N_{0}=\frac{1}{8m\omega_{H}D_{0}\tau}$---is the maximal number of Landau
levels determined by the cutoff.

With accuracy sufficient for the use in Eq.(\ref{Hc}), the solution of Eqs.
(\ref{sisdif1}) for diffusion coefficient in Cooper channel in small magnetic
fields when $\omega_{H}\ll\frac{\lambda e^{-1/\lambda}}{\tau}$, can be written
in the following form [3]:
\begin{equation}
d_{2}=
\left\{\begin{array}{ll}
1 & \mbox{ при } \omega\gg\frac{e^{-1/\lambda}}{2\tau} \\
2\omega\tau e^{1/\lambda} & \mbox{ при }\omega\ll\frac{e^{-1/\lambda}}{2\tau}
\end{array}
\right.
\label{d2litl}
\end{equation}
and in fact we can neglect the magnetic field influence upon diffusion.

It is easy to see that the anomalies of the upper critical field due to the
frequency dependence of diffusion coefficient will appear only for temperatures
$T\ll\frac{e^{-1/\lambda}}{\tau}$ [3]. For higher temperatures we obtain the
usual behavior of "dirty" superconductors. Superconductivity survive in a
system with finite localization length if the following inequality holds
$T_{c}\gg \lambda\frac{e^{-1/\lambda}}{\tau}$ [3], which is equivalent
to the well known criteria [9] of the smallness of Cooper pair size compared
with localization length. This latter length is exponentially large in two-
dimensional systems with small disorder ($\lambda\ll 1$). The most
interesting (for our aims) limit of relatively strong disorder is defined by
$T_{c}\ll \frac{e^{-1/\lambda}}{\tau}$, so that in fact we are dealing with
pretty narrow region of $\lambda$'s when
$\lambda\frac{e^{-1/\lambda}}{\tau}\ll T_{c}\ll \frac{e^{-1/\lambda}}{\tau}$.
In this case the upper critical field is defined by the equation
($\gamma=1.781$)[3]:
\begin{equation}
ln\left(\frac{\gamma}{2\pi}\frac{e^{-1/\lambda}}{\tau
T}\right)=(1+4\pi\frac{D_{0}}{\Phi_{0}}\frac{\tau
H_{c2}}{e^{-1/\lambda}})ln\left(\frac{\gamma}{2\pi}\frac{e^{-1/\lambda}}{\tau
T_{c}}(1+4\pi\frac{D_{0}}{\Phi_{0}}\frac{\tau H_{c2}}{e^{-1/\lambda}})\right)
\label{Hc3}
\end{equation}
from which we can directly obtain the $T(H_{c2})$---dependence.
The appropriate behavior of the upper critical field for two sets of parameters
is shown in Fig.1. The curve of $H_{c2}(T)$ demonstrates negative curvature
and $H_{c2}$ diverges for $T\rightarrow 0$. This weak (logarithmic) divergence
is connected with our neglect of the magnetic field influence upon diffusion
[3]. Taking this influence into account we can suppress this divergence of
$H_{c2}$ as $T\rightarrow 0$ and we obtain:
\begin{equation}
H_{c2}(T=0)=\frac{\gamma}{2\pi}\frac{\Phi_{0}}{D_{0}} \frac{1}{\tau}
\label{Hc(0)H}
\end{equation}
This is the main effect of broken time invariance and it is clear that it is
important only for extremely low temperatures [3]. In the following we neglect
it.

For the quasi-two-dimensional case  on the dielectric side of Anderson's
transition, but not too very close to it, the behavior of diffusion coefficient
is quite close to that of purely two-dimensional case, so that the upper
critical field can be analyzed within two-dimensional approach. Close to the
transition (e.g. over interplane transfer integral) both for metallic and
insulating sides and for parameters satisfying the inequality
$\lambda\frac{e^{-1/\lambda}}{\tau}\ll T_{c}\ll \frac{e^{-1/\lambda}}{\tau}$,
the temperature dependence of $H_{c2}$ is in fact again very close to
those in purely two-dimensional case considered above [3]. Some deviations
appear only in a very narrow region of very low temperatures [3]. The details
can be found in Ref. [3].

These anomalies of $H_{c2}$---behavior are always due to the appropriate
anomalies of frequency dependence of the generalized diffusion coefficient
close to the Anderson transition and in this sense just reflect the change in
the nature electronic states in the region of metal-insulator transition.

In Fig.1 we also show the experimental data for $H_{c2}$ from Ref.[1].
Theoretical curve (1) is given for the parameters which lead to rather good
agreement with experiment in the low temperature region. The curve (2)
corresponds to parameters giving good agreement in a wide temperature region
except the lowest temperatures. The cyclotron mass $m$ was always assumed to be
equal that of the free electron. In general we observe satisfactory agreement
between theory and experiment. Unfortunately, the values of the ratio
$\frac{e^{-1/\lambda}}{T_{c}\tau}$ for the second curve, while corresponding to
quite reasonable values of $\lambda$, lead to nonrealistic (too small) values
of $T_{c}\tau$, which are rather doubtful for the system with relatively
high $T_{c}$. For the first curve situation is much better though the size
of electron damping on the scale of $T_{c}$ is still very large which
corresponds to strong disorder. Note however, that the detailed discussion of
these parameters is actually impossible without the knowledge of additional
characteristics of the films studied in Ref. [1]. In particular it is quite
interesting to have an independent estimate of $\lambda$. We also want to
stress relatively approximate nature of these parameters due to our two-
dimensional idealization. More serious comparison should be done using the
expressions of Ref. [3] for the quasi-two-dimensional case, which again
requires the additional information on the system, in particular, the data on
the anisotropy of electronic properties.

In our opinion the relatively good agreement of experimental data of Ref. [1]
with theoretical dependences obtained for the two-dimensional (quasi-two-
dimensional) case of disordered system with Anderson localization
illustrates the importance of localization effects for the physics of high-
temperature superconductors [10].

\vskip 0.5cm
The authors are grateful to Dr. I.Bozovic for sending the preprint of
Ref.[1] before its publication. This work is supported by Russian Foundation
for Fundamental Research under the grant $N^{o}$ 93-02-2066.

\newpage
\setlength{\textwidth}{16.5cm}

\newpage
\setlength{\textwidth}{15.0cm}
\begin{center}
{\bf Figure Captions:}
\end{center}
\vskip 0.5cm

Fig.1 Temperature dependence of the upper critical field:

Theoretical curve (1) is given for the case of
$\frac{e^{-1/\lambda}}{T_{c}\tau}=2$, $\lambda=0.18$, while the curve

(2) is for $\frac{e^{-1/\lambda}}{T_{c}\tau}=20$, $\lambda=0.032$.

Squares---experimental data of Ref. [1].

\end{document}